# Model-free approach to the interpretation of restricted and anisotropic self-diffusion in magnetic resonance of biological tissues

Omar Narvaez[1], Maxime Yon[2], Hong Jiang[2], Diana Bernin[3], Eva Forssell-Aronsson[4,5], Alejandra Sierra[1], Daniel Topgaard[2,*]

[1]*A.I.Virtanen Institute for Molecular Sciences, University of Eastern Finland, Kuopio, Finland.*
[2]*Department of Chemistry, Lund University, Lund, Sweden.*
[3]*Department of Chemical Engineering, Chalmers University of Technology, Gothenburg, Sweden.*
[4]*Department of Medical Radiation Sciences, University of Gothenburg, Gothenburg, Sweden.*
[5]*Medical Physics and Biomedical Engineering, Sahlgrenska University Hospital, Gothenburg, Sweden.*
*email: daniel.topgaard@fkem1.lu.se

Magnetic resonance imaging (MRI) is the method of choice for noninvasive studies of micrometer-scale structures in biological tissues via their effects on the time/frequency-dependent ("restricted) and anisotropic self-diffusion of water. Traditional MRI relies on pulsed magnetic field gradients to encode the signal with information about translational motion in the direction of the gradient, which convolves fundamentally different aspects—such as bulk diffusivity, restriction, anisotropy, and flow—into a single effective observable lacking specificity to distinguish between biologically plausible microstructural scenarios. To overcome this limitation, we introduce a formal analogy between measuring rotational correlation functions and interaction tensor anisotropies in nuclear magnetic resonance (NMR) spectroscopy and investigating translational motion in MRI, which we utilize to convert data acquisition and analysis strategies from NMR of rotational dynamics in macromolecules to MRI of diffusion in biological tissues, yielding model-independent quantitative metrics reporting on relevant microstructural properties with unprecedented specificity. Our model-free approach advances the state-of-the-art in microstructural MRI, thereby enabling new applications to complex multi-component tissues prevalent in both tumors and healthy brain.

Subject Areas: Chemical Physics, Medical Physics, Physical Chemistry

## I. INTRODUCTION

Nuclear magnetic resonance (NMR) and magnetic resonance imaging (MRI) offer noninvasive characterization of cellular-level structures in intact biological tissues by employing time-varying magnetic field gradients to monitor the micrometer-scale translational motion of water molecules [1,2] and, by inference, their interactions with cell membranes and macromolecules [3]. While the use of diffusion MRI is in current clinical practice limited to rather basic measurements of diffusion-weighted images and apparent diffusion coefficients [4] to detect and grade ischemic stroke [5] and tumors [6], there is a recent trend of applying increasingly advanced motion-encoding gradients to isolate specific aspects such as diffusion anisotropy [7] and time/frequency-dependence [8], the latter traditionally known as "restricted diffusion" [9,10]. Despite the developments of specific encoding strategies and numerous examples of promising applications in clinical research [11], data analysis and interpretation remain a challenge—in particular for heterogeneous tissues where each imaging voxel contains multiple water populations, or "pools", with distinct anisotropy and restriction properties [12,13].

The details of translational motion are described with the velocity autocorrelation function [14] and its Fourier transform, the tensor-valued diffusion spectrum $\mathbf{D}(\omega)$ [15], which may be interrogated by applying modulated gradients with encoding spectra $\mathbf{b}(\omega)$ having peaks at selected frequencies $\omega$ [16]. The $\omega$-dependence of $\mathbf{D}(\omega)$ has been derived for simple pore shapes such as parallel planes, cylinders, and spheres [17], as well as for more elaborate geometries including the random permeable membranes model [18]. The experimentally accessible range of $\omega$ is determined by the performance of the gradient hardware and is in practice often limited to the rather narrow ranges 10-50 Hz for clinical and 1-1000 Hz for pre-clinical MRI systems, thus making it difficult to distinguish between different candidate models from the observed $\omega$-dependence alone [19]. The problem becomes even more severe when studying heterogeneous tissues with separate water pools potentially having different $\omega$-dependence.

Related ambiguities occur in NMR relaxation studies of molecular reorientation where multiple, equally plausible, models may be consistent with the experimental observations. In this area, dynamics is quantified with orientation autocorrelation functions, often assumed to be multiexponential, and the corresponding spectral densities $J(\omega)$, which are probed by measuring relaxation rate constants determined by the values of $J(\omega)$ at sets of discrete frequencies given by the applied static and radiofrequency magnetic fields and the gyromagnetic ratios of the involved atomic nuclei [20]. In a highly influential paper, Lipari and Szabo introduced a "model-free approach" to convert relaxation rates measured for macromolecules in solution into a few unique dynamics parameters consistent with more sophisticated models [21]. These ideas were recently generalized by the concept of dynamics detectors [22-24] where an approximation of $J(\omega)$ as a nonparametric distribution of Lorentzians enables conversion of a discrete set of relaxation observables to the average amplitudes of motion within specific ranges of rotational correlation times without having to invoke an explicit motional model. Independently of the information about dynamics, resolution of different atomic sites is in NMR spectroscopy achieved by multidimensional separation and correlation of chemical shifts [25], including isotropic-anisotropic correlations in solid-state NMR [26].

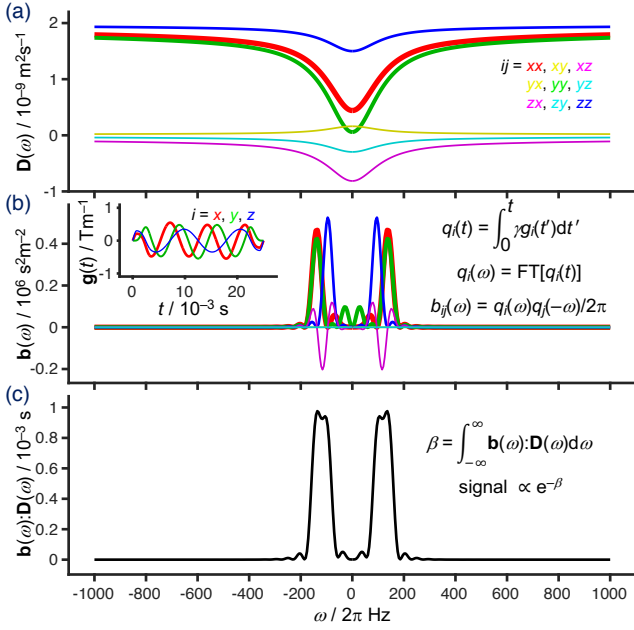

Fig. 1. Principles for $\mathbf{b}(\omega)$-encoding of the magnetic resonance signal with information about restricted and anisotropic diffusion. (a) Tensor-valued diffusion spectrum $\mathbf{D}(\omega)$ for a liquid with bulk diffusivity $D_0 = 2\cdot10^{-9}$ m$^2$s$^{-1}$ confined in a cylindrical compartment with radius $r = 3$ μm and orientation in the lab frame given by the polar and azimuthal angles $\theta = -30°$ and $\phi = 20°$ (see Eqs. (20)-(25) in methods). Color coding of the elements $D_{ij}(\omega)$ is given in the legend to the right. (b) Tensor-valued encoding spectrum $\mathbf{b}(\omega)$ corresponding to the time-dependent gradient vector $\mathbf{g}(t)$ shown in the inset to the left. The relations between the Cartesian components $g_i(t)$ and elements $b_{ij}(\omega)$ are given in the equations to the right, where $\gamma$ is the gyromagnetic ratio, $\mathbf{q}(t)$ is the time-dependent dephasing vector, and FT[$x$] denotes a Fourier transformation. (c) Generalized scalar product $\mathbf{b}(\omega):\mathbf{D}(\omega)$, which upon integration over frequencies $\omega$ gives the attenuation factor $\beta$ and signal according to the equations to the right.

Building on these insights and after identification of some key formal analogies revealing that both the $\omega$-dependence in relaxation NMR and tensorial aspect in solid-state NMR are captured in the composite acquisition variable $\mathbf{b}(\omega)$ of diffusion MRI (see methods for details), we introduce a model-free approach to quantify restricted and anisotropic diffusion of water in heterogeneous biological tissues in terms of nonparametric distributions of tensor-valued Lorentzians. In the limit of low frequencies, these novel "$\mathbf{D}(\omega)$-distributions" are equivalent to the ($\omega$-inde-pendent) discrete [27] or continuous [28,29] diffusion tensor distributions that are ubiquitous for analysis of diffusion anisotropy in heterogeneous brain tissues. Signal encoding using the principles of isotropic-anisotropic shift correlation in solid-state NMR [30] and data inversion with Monte Carlo methods from Laplace NMR [31] have recently enabled estimation of nonparametric tensor distributions in which distinct water pools may be identified as clusters of components in an analysis space spanned by the isotropic, anisotropic, and orientation dimensions [32,33]. Here we augment the previous data acquisition scheme with exploration of the $\omega$-dimension of $\mathbf{b}(\omega)$ to allow estimation of the restriction properties for each of the water pools resolved in other dimensions.

The potential of the new method is demonstrated on MRI phantoms with multiple well-defined water pools, *ex vivo* rat brain, and excised tissue from a xenograft model of neuroblastoma [34]. Encouraged by the recent profusion of *in vivo* human studies using nearly identical MRI pulse sequences to explore either the spectral [35-41] or tensorial [42-47] aspects of the encoding, we envision that the unification of the traditionally separate encoding strategies into a common framework, enabled by our model-free approach to data analysis, will catalyze the design of more informative and time-efficient data acquisition protocols for clinical research studies of tissue microstructure in health and disease.

## II. RESULTS

The principles for using time-varying magnetic field gradients $\mathbf{g}(t)$ to investigate tensor-valued diffusion spectra $\mathbf{D}(\omega)$ are illustrated in Fig. 1. For a closed compartment, the $\omega$-dependence of each of the eigenvalues of $\mathbf{D}(\omega)$ can be written as a sum of Lorentzians with varying widths and amplitudes [17] (see methods for details). The gradients define an encoding spectrum $\mathbf{b}(\omega)$ which determines the signal attenuation via the integral of the generalized scalar product $\mathbf{b}(\omega):\mathbf{D}(\omega)$ over $\omega$, implying that $\mathbf{D}(\omega)$ can be reconstructed from a series of measurement in which the frequency content of $\mathbf{b}(\omega)$ is varied [48]. Resolution of water populations with distinct $\mathbf{D}(\omega)$ requires measurements with varying anisotropy of $\mathbf{b}(\omega)$ [7,32]—preferably at each value of $\omega$. Ideal measurements would be performed with $\mathbf{b}(\omega)$ having varying anisotropy and being finite at only a single frequency, thus allowing $\mathbf{D}(\omega)$ to mapped out frequency by frequency. In practice, $\mathbf{b}(\omega)$ invariably comprises a range of frequencies as illustrated in Fig. 1. Gradient waveforms derived from the analogy between sample spinning in solid-state NMR and $q$-vector trajectories in diffusion NMR [49], in this specific case the double rotation technique [50,51], allow generation of encoding spectra $\mathbf{b}(\omega)$ with reasonably narrow spectral content in multiple dimensions. Similarly to the widely used cos-modulated oscillating gradients [52], the characteristic frequencies are adjustable by the number of gradient amplitude periods within the duration of the waveform. Despite having more well-defined spectral content than earlier incarnations of tensor-valued encoding [53], the spread over frequencies means that each individual measurement contains entangled information on restriction and anisotropy.

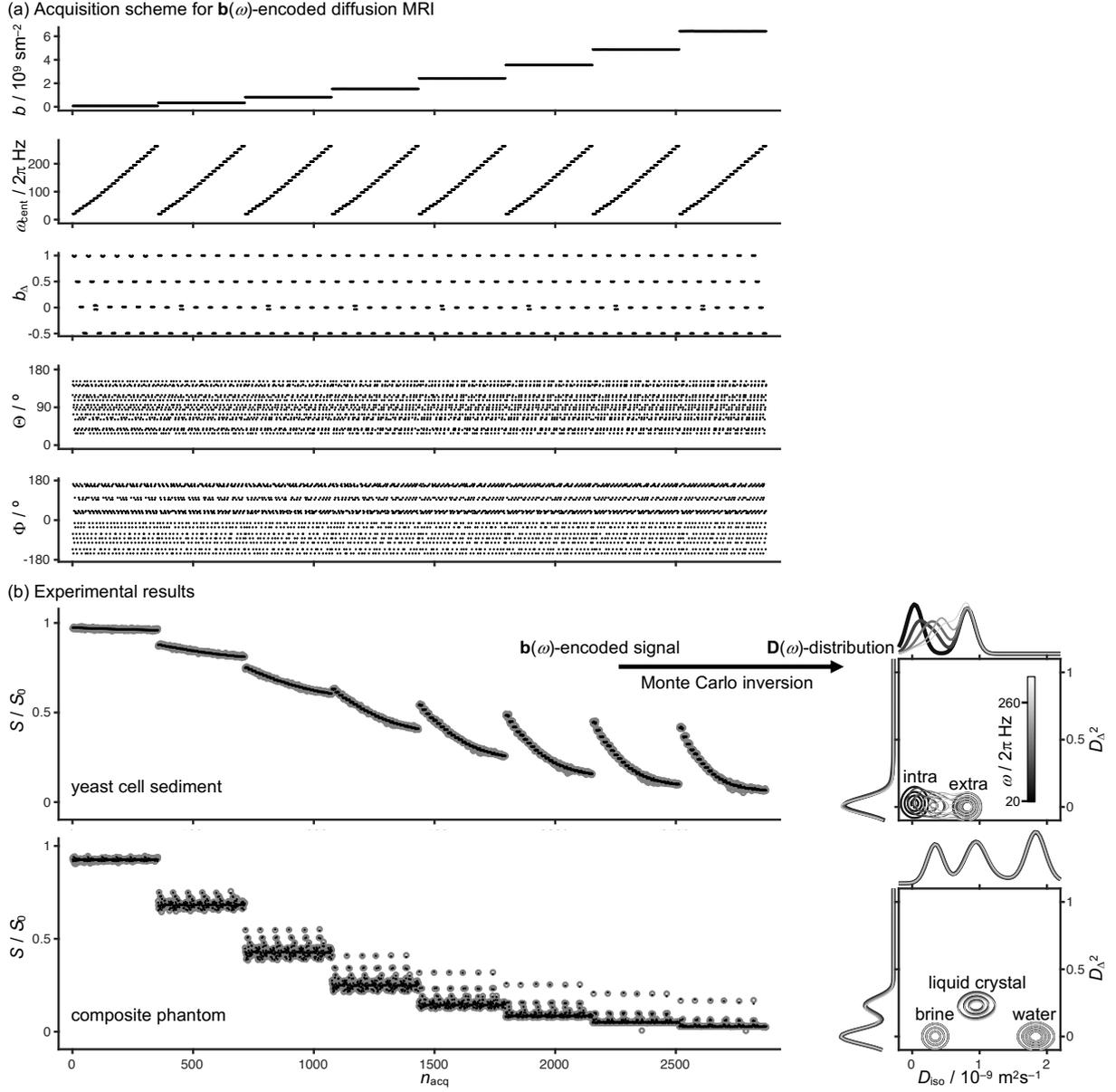

Fig. 2. Comprehensive acquisition scheme and experimental results for $\mathbf{b}(\omega)$-encoded diffusion MRI. (a) Magnitude $b$, centroid frequency $\omega_{\mathrm{cent}}$, normalized anisotropy $b_\Delta$, and orientation $(\Theta,\Phi)$ of the tensor-valued encoding spectrum $\mathbf{b}(\omega)$ vs. the acquisition number $n_{\mathrm{acq}}$ with maximum value 2880. Diffusion encoding was performed with pairs of gradient waveforms of the type shown in Fig. 1 with 25 ms duration and 3 Tm$^{-1}$ maximum amplitude. (b) Experimental data (circles: measured, points: back-calculated from the $\mathbf{D}(\omega)$-distributions) obtained at 11.7 T on a yeast cell sediment and a composite phantom comprising an assembly of glass tubes with pure water, an aqueous solution saturated with magnesium nitrate (brine), and a lamellar liquid crystal of water, sodium decanoate, and decanol. Monte Carlo inversions of the $\mathbf{b}(\omega)$-encoded signals yield $\mathbf{D}(\omega)$-distributions shown in the panels to the right as projections onto the 2D plane and 1D axes of the isotropic diffusivity $D_{\mathrm{iso}}$ and squared normalized anisotropy $D_\Delta^2$ for five values of $\omega$ (indicated with linear gray scale of contour lines). The intracellular water in the yeast is restricted ($\omega$-dependent) while the four other water pools are Gaussian ($\omega$-independent) within the investigated range of $\omega/2\pi$ from 20 to 260 Hz.

As a solution to this problem, we here propose to estimate nonparametric $\mathbf{D}(\omega)$-distributions by global inversion of data acquired as a function $\mathbf{b}(\omega)$ with varying magnitude $b$, spectral content summarized by the centroid frequency $\omega_{\mathrm{cent}}$ [36], normalized anisotropy $b_\Delta$ [54], and orientation $(\Theta,\Phi)$ (see Eqs. (4)-(8) in methods). The requirement of acquiring data at unique values of $\omega$ may be relaxed by sampling a range of spectral contents and invoking physically reasonable constraints on the components of the $\mathbf{D}(\omega)$-distributions—here by assuming that the $\omega$-dependence of tensor eigenvalues is Lorentzian and the tensor shapes are axisymmetric [54] (see methods for justification of the assumptions). Under these constraints, each component of the distribution is described by its weight $w$, low-$\omega$ axial and radial diffusivities $D_\mathrm{A}$ and $D_\mathrm{R}$, polar and azimuthal angles $\theta$ and $\phi$, high-$\omega$ isotropic diffusivity $D_0$, and axial and radial transition rates $\Gamma_\mathrm{A}$ and $\Gamma_\mathrm{R}$. The $\mathbf{b}(\omega)$-encoded signal for a single component is proportional to exp(–$\beta$) where the attenuation factor $\beta$ is obtained by numerical integration of the generalized scalar product $\mathbf{b}(\omega):\mathbf{D}(\omega)$ over frequencies [48] and the relation between $\mathbf{D}(\omega)$ and the parameters $[D_A,D_R,\theta,\phi,D_0,\Gamma_A,\Gamma_R]$ are given in Eqs. (19) and (20) in methods. The new approach may be recognized as an extension of previous diffusion tensor distributions [27-29] with Lorentzian $\omega$-dependence of the tensor eigenvalues, corresponding to exponential velocity autocorrelation functions [14] with decay rate $\Gamma$. Following earlier works [55], Monte Carlo inversion [31] is used to estimate ensembles of distributions consistent with the measured data. For visualization, the $\mathbf{D}(\omega)$-distributions in the primary analysis space $[D_A,D_R,\theta,\phi,D_0,\Gamma_A,\Gamma_R]$ are evaluated at selected values of $\omega$, giving $[D_A(\omega),D_R(\omega),\theta,\phi]$, and projected onto the dimensions of isotropic diffusivity $D_{iso}(\omega)$ and squared normalized anisotropy $D_\Delta(\omega)^2$ [56], as well as the lab-frame diagonal values $D_{xx}(\omega)$, $D_{yy}(\omega)$, and $D_{zz}(\omega)$. Although the $\mathbf{D}(\omega)$-distributions are defined for all values of $\omega$, only the rather modest range between the minimum and maximum values of $\omega_{cent}$ have been properly investigated in the encoding process and are meaningful to interpret. For generating parameter maps, the rich information in the $\mathbf{D}(\omega)$-distributions is condensed into means E[$x$], variances V[$x$], and covariances C[$x,y$] over relevant dimensions and subdivisions ("bins") of the distribution space [33].

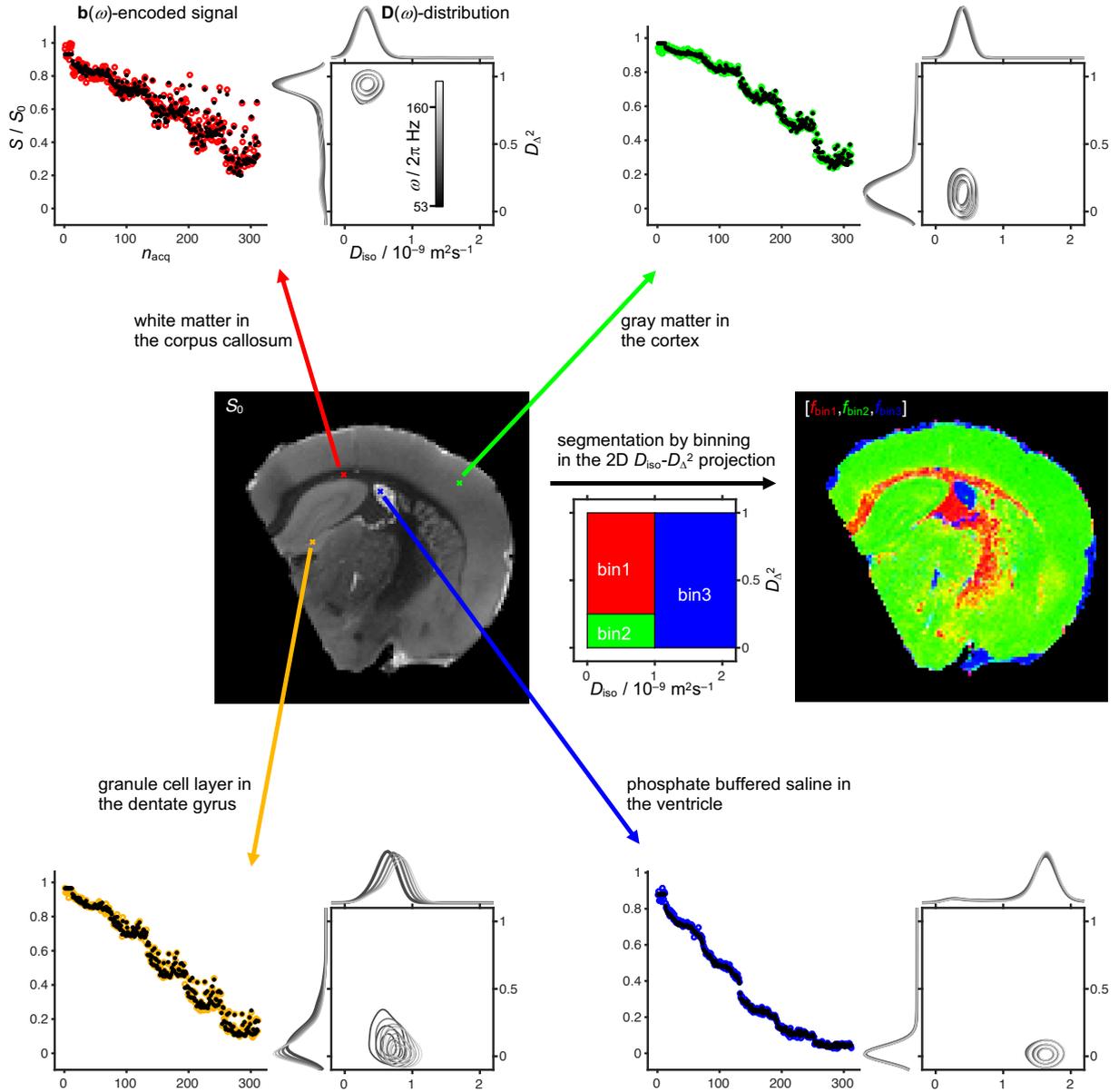

Fig. 3. $\mathbf{D}(\omega)$-distributions for selected voxels in *ex vivo* rat brain. The figures show $\mathbf{b}(\omega)$-encoded signals and corresponding $\mathbf{D}(\omega)$-distributions for the four voxels indicated with crosses in the $T_2$-weighted image $S_0$. The acquisition scheme is a 312-point abbreviated version of the 2880-point comprehensive one in Fig. 2 and limited to the range of $b$-values up to $3.6 \cdot 10^9$ sm$^{-2}$ and $\omega_{cent}/2\pi$ from 53 to 160 Hz. The results for the individual voxels at $\omega/2\pi = 53$ Hz guide the division of the 2D $D_{iso}$-$D_\Delta^2$ projection into three bins—nominally specific for white matter, gray matter, and phosphate buffered saline—for the purpose of image segmentation by coding the per-bin signal fractions $f_{bin1}$, $f_{bin2}$, and $f_{bin3}$ into RGB color and extraction of bin-specific diffusion metrics. The voxels from the granule cell layer in the dentate gyrus and white matter show the hallmarks of restriction ($\omega$-dependence) and anisotropy ($D_\Delta^2 \approx 1$), respectively.

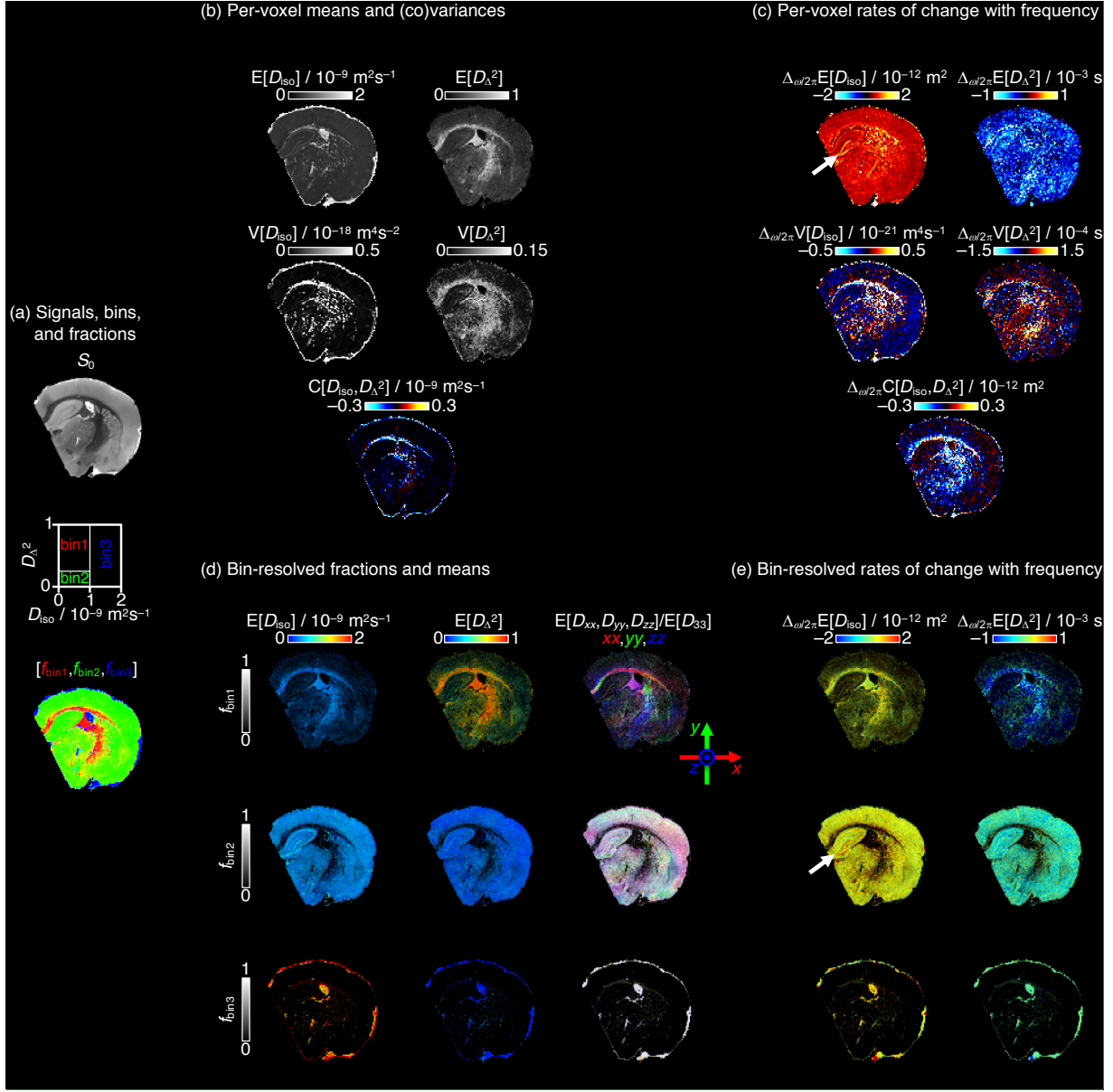

Fig. 4. Parameter maps derived from the per-voxel $\mathbf{D}(\omega)$-distributions for *ex vivo* rat brain. (a) Synthesized $T_2$-weighted image $S_0$, definition of bins in the 2D $D_{iso}$-$D_\Delta^2$ projection, and map of per-bin signal fractions [$f_{bin1}, f_{bin2}, f_{bin3}$] coded into RGB color. (b) Per-voxel statistical descriptors E[$x$], V[$x$], and C[$x,y$] over the $D_{iso}$ and $D_\Delta^2$ dimensions of the $\mathbf{D}(\omega)$-distributions evaluated at a selected frequency $\omega/2\pi = 53$ Hz. (c) Rates of change with frequency, $\Delta_{\omega/2\pi}$, of the per-voxel metrics highlighting areas with effects of restricted diffusion. (d) Bin-resolved signal fractions and means E[$x$] of the diffusion metrics at $\omega/2\pi = 53$ Hz coded into image brightness (vertical brightness bars) and blue-green-red color scale (horizontal color bars). Color-coding of orientation derives from the lab-frame (shown with red, green, and blue arrows) diagonal values [$D_{xx}, D_{yy}, D_{zz}$] normalized by the maximum eigenvalue $D_{33}$. (e) Bin-resolved rates of change with frequency. The white arrows in panels (c) and (e) indicate elevated values of $\Delta_{\omega/2\pi}$E[$D_{iso}$] for the granule cell layer in the dentate gyrus.

Experimental demonstration of the approach is given in Fig. 2 for two samples with well-defined and previously investigated restriction and anisotropy properties, namely a yeast cell sediment [57] and an assembly of glass tubes with pure water [58], saturated salt solution [59], and lamellar liquid crystal [60]. In the case of isotropic Gaussian diffusion, the signal attenuation is completely determined by the value of $b$ and independent of all other variables $\omega_{cent}$, $b_\Delta$, $\Theta$, and $\Phi$. The observed sensitivity to $\omega_{cent}/2\pi$ in the investigated range 20-260 Hz for the yeast cell sediment indicates restriction in micrometer-scale compartments, while the dependence on $b_\Delta$ for the composite phantom reveals anisotropy. These qualitative observations of restriction and anisotropy from the raw signal data are filled in with more details by the obtained $\mathbf{D}(\omega)$-distributions: The yeast sample comprises two isotropic ($D_\Delta^2 = 0$) pools, one Gaussian ($\omega$-independent) and one restricted ($\omega$-dependent) originating from, respectively, the extra- and intracellular spaces separated by the virtually impermeable plasma membranes [61]. The composite phantom yields three Gaussian pools, one of which being anisotropic with a value $D_\Delta^2 = 0.25$ consistent with the essentially two-dimensional diffusion of water confined to the nanometer-scale gaps between the planar detergent bilayers in a lamellar liquid crystal [62]. With values of $\omega_{cent}/2\pi$ on the scale of $10^9$ Hz, it would be possible to observe effects of restricted diffusion across these gaps. Correspondingly, values of $\omega_{cent}/2\pi$ approaching $10^{12}$ Hz would allow investigating the regime of ballistic motion of the individual water molecules [63]. These high-$\omega$ regimes are however far beyond the range accessible with MRI methods based on magnetic field gradients.

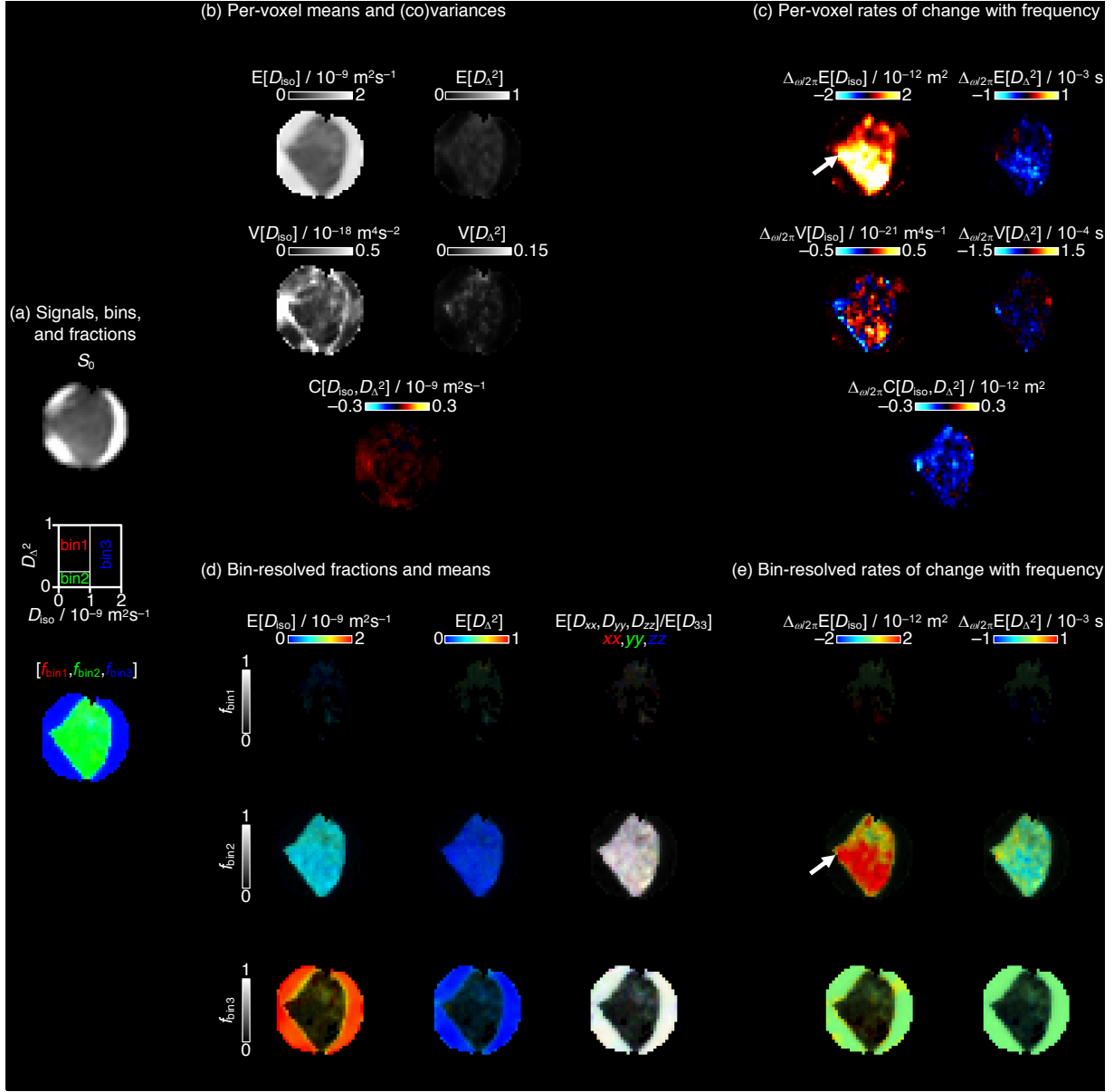

Fig. 5 Parameter maps for part of an excised tumor immersed in aqueous formaldehyde solution. See caption of Fig. 4 for detailed explanation of the panels. The acquisition scheme is a 480-point abbreviated version of the 2880-point comprehensive one in Fig. 2 and limited to the range $\omega_{cent}/2\pi$ from 44 to 140 Hz using $b$-values up to $11 \cdot 10^9$ sm$^{-2}$. The arrows in panels (c) and (e) show tumor areas with pronounced effects of restricted diffusion.

Fig. 3 shows data for a few representative voxels in an *ex vivo* rat brain. The $\mathbf{D}(\omega)$-distributions for voxels in pure white matter (WM), gray matter (GM), and phosphate buffered saline (PBS) in the ventricles are qualitatively consistent with earlier *in vivo* mouse results [64] (WM: low $D_{iso}$ and high $D_\Delta^2$, GM: low $D_{iso}$ and low $D_\Delta^2$, and PBS: high $D_{iso}$ and low $D_\Delta^2$) with only barely detectable $\omega$-dependence in the investigated range from 53 to 160 Hz. The voxel in the granule cell layer in the dentate gyrus gives a $\mathbf{D}(\omega)$-distribution resembling the one from GM, but with more pronounced $\omega$-dependence in agreement with earlier observations using oscillating gradient encoding [65]. The $\omega$-dependence is consistent with granule cell dimensions on the 10 μm scale as seen in histology [66]. The $\mathbf{D}(\omega)$-distributions for WM, GM, and PBS guide the definition of three bins in the 2D $D_{iso}$-$D_\Delta^2$ projection to generate maps of nominally tissue type-specific per-bin signal fractions and diffusion metrics. Fig. 4 compiles maps of per-voxel and bin-resolved statistical descriptors E[$D_{iso}$], E[$D_\Delta^2$], V[$D_{iso}$], V[$D_\Delta^2$], and C[$D_{iso}$,$D_\Delta^2$], typically associated with tensor-valued encoding [11,33], and rates of change of the diffusion metrics with frequency, for instance $\Delta_{\omega/2\pi}$E[$D_{iso}$] often used to display results from oscillating gradient encoding [36,65,67], as well as novel metrics correlating information about restriction and anisotropy. Of special note in this latter category is the separation of high- and low-$D_\Delta^2$ components with similar $D_{iso}$ (bin1 and bin2), and the association of the effects of restricted diffusion to the low-$D_\Delta^2$ component (bin2) in Fig. 4(e).

The maps of the excised tumor in Fig. 5 feature an extended area with pronounced effects of restricted diffusion (high $\Delta_{\omega/2\pi}E[D_{iso}]$) as well as regions with high $V[D_{iso}]$ resulting from the co-existence of low- and high-$D_{iso}$ water pools within the same imaging voxel, tentatively originating from dense tumor and formaldehyde solution or tissues with degraded cell membranes. These maps may permit non-invasive resolution of apoptotic, necrotic, and viable tumor tissues having distinctly different cell densities, sizes, shapes, and membrane properties.

### III. DISCUSSION

The set of experimental data demonstrate that our proposed model-free approach enables estimation of quantitative metrics related to restricted and anisotropic diffusion within a single set of measurements, thereby merging oscillating gradient and tensor-valued encoding into a common experimental and analysis framework. Although the complete tensor-valued encoding spectra $\mathbf{b}(\omega)$ are used in the data inversion, the acquisition scheme in Fig. 2 is reported in terms of the five variables $b$, $\Theta$, $\Phi$, $\omega_{cent}$, and $b_\Delta$ (see definitions in Eqs. (4)-(8) in methods). These variables provide a convenient short-hand notation of the investigated diffusion properties and are all familiar from the literature—the first three being the $b$-value and -vector of diffusion tensor imaging [68,69], the fourth the characteristic encoding frequency of oscillating gradients [16,36,52], and the fifth the normalized anisotropy of tensor-valued encoding [54]. At high $\omega_{cent}$ and $b_\Delta = 0$, the signal as a function of $b$ depends solely on the distribution of high-$\omega$ isotropic diffusivities. Decreasing $\omega_{cent}$ and using non-zero values of $b_\Delta$ bring in the effects of, respectively, restriction and anisotropy. At the lowest values of $\omega_{cent}$ and $b_\Delta = 1$, these effects are maximized and the variation of signal with $(\Theta,\Phi)$ gives information about the orientations of anisotropic compartments. Conventional diffusion MRI [68,69] is performed in this latter limit and is sensitive to all of the microstructural properties without being able to resolve their individual contributions [51]. Sampling of the multidimensional space spanned by all five effective acquisition variables, on the other hand, allows retrieval of the corresponding multidimensional $\mathbf{D}(\omega)$-distributions with information about diffusivity, orientation, restriction, anisotropy, and their correlations.

The quality of the fits in Fig. 2 and Fig. 4 illustrate that our proposed signal expression, being the sum of contributions from components with $\mathbf{D}(\omega)$ approximated as tensor-valued Lorentzians (see Eqs. (17) and (19) in methods), is sufficiently flexible to capture all relevant signal modulations over exhaustive ranges of acquisition variables for the investigated samples selected for their well-known effects of restrictions and anisotropy. Correspondingly, the obtained $\mathbf{D}(\omega)$-distributions in Fig. 2 and Fig. 3 and derived parameter maps in Fig. 4 and Fig. 5 are all consistent with the design of the phantoms and previous results in the literature, showing that the good fits do not come at the expense of excessive overfitting that would lead to spurious peaks in the distributions and noisy parameter maps.

Despite the successful data fitting and reproduction of expected results, we emphasize that the proposed approximation is merely a convenient mathematical representation that derives from the corresponding expressions for restricted diffusion in closed compartments (see Eq. (22) in methods). In our attempt to find an acceptable compromise between physical correctness, mathematical convenience, and utility for solving scientific question without tempting the user to overinterpretation, we have drawn inspiration from the history of the development of methods to extract information about molecular reorientation from NMR relaxation data—more specifically the evolution from the Lipari-Szabo model-free approach [21] to the recent dynamics detectors [22-24]. The rotational motion of, for instance, a carbon-hydrogen bond in a protein is governed by the forces acting between the atoms in the bond and other atoms in the same and neighboring molecules. The intricate details of these forces may lead to rather elaborate dynamics that for conceptual simplicity often is modeled as a set of statistically independent rotational diffusion processes with correlation functions written as a product of exponentials, which by assuming separation of time scales may be approximated as a *sum* of exponentials that, via Fourier transformation, gives spectral densities as a sum of Lorentzians. The mathematical convenience of this latter functional form, combined with its excellent ability to fit experimental data, has made it the point of departure in most contemporary works on rotational dynamics and NMR relaxation. While many studies still discuss whether two, three, or a continuous distribution of Lorentzian components are required to fit the experimental data, the latest key development is the dynamics detectors [22-24] that have enabled extraction of information about the total amplitudes of motion within a few broad and partially overlapping ranges of correlation times. Despite yielding information on a coarser scale than traditional approaches, the responses of the dynamics detectors are independent of the exact details of the underlying distributions and are, consequently, less conducive to overinterpretation. These lessons from NMR relaxation are closely mirrored in our proposed approach with tensor-valued Lorentzian $\mathbf{D}(\omega)$-components, ensembles of nonparametric $\mathbf{D}(\omega)$-distribution estimated from the signal data via Monte Carlo inversion, and, finally, extraction of coarser-level projections and quantitative metrics that report on the relevant properties without providing ambiguous details that are not necessarily required by the input data.

With the comprehensive 2880-point acquisition scheme in Fig. 2, the presence of restriction and anisotropy can be deduced by simple visual inspection of the signal intensities as a function of the acquisition variables—especially $\omega_{cent}$ and $b_\Delta$ for a given value of $b$—and quantified from the 2D $D_{iso}$-$D_\Delta^2$ projections of the obtained $\mathbf{D}(\omega)$-distributions. Ad-mittedly, the data in Fig. 2 were acquired under exceptionally favorable circumstances, using 3 Tm$^{-1}$ gradient hardware and samples with sufficiently large values of the transverse relaxation time $T_2$ to allow for in total 0.050 s of diffusion-encoding gradients and the broad range 20-260 Hz of $\omega_\mathrm{cent}/2\pi$ even at the highest $b$-value 6.4·10$^9$ sm$^{-2}$. Conversely, the data in Fig. 3 to Fig. 5 represent more realistic conditions with short-$T_2$ fixated tissues and abbreviated acquisition schemes comprising only 312 or 480 data points over the limited ranges 53-160 or 44-140 Hz. Despite these limitations, the data in Fig. 4 reproduce earlier findings on both restrictions [65] and anisotropy [64] in rodent brain, as well as bring novel information on the correlations between the properties. The number of acquisitions is comparable to the 10-min and 300-point schemes used in early clinical implementations of tensor-valued encoding for studies of brain tumors [70], later to be truncated and optimized for 3-min measurements consistent with applications in clinical practice [44], thus indicating the potential for implementation of our proposed method for both clinical and pre-clinical research studies—initially maybe by simply interleaving the very latest protocols for oscillating gradient [36] and tensor-valued encoding [47] using identical pulse sequences and imaging settings.

Table 1. MRI acquisition and processing parameters

| | phantoms | *ex vivo* rat brain | excised tumor |
|---|---|---|---|
| spectrometer | Avance-Neo | Avance-III HD | Avance-III HD |
| magnetic field / T | 11.7 | 11.7 | 14.0 |
| acquisition software | TopSpin 4.0.7 | ParaVision 6.0.1 | TopSpin 3.5.6 |
| image read-out | RARE | MSME | RARE |
| acquisition resolution / mm$^3$ | 0.15×0.3×1 | 0.09×0.09×0.09 | 0.15×0.3×1 |
| matrix size | 32×16×1 | 111×111×10 | 32×16×1 |
| diffusion gradient duration / ms | 25 | 8 | 10 |
| max $b$-value / $10^9$ sm$^{-2}$ | 6.4 | 3.5 | 11 |
| centroid frequency $\omega_\mathrm{cent}/2\pi$ Hz | 20-260 | 53-160 | 44-140 |
| normalized anisotropy $b_\Delta$ | –0.5, 0, 0.5, 1 | –0.5, 0, 0.5, 1 | –0.5, 0, 0.5, 1 |
| # directions | 15 | 11 | 15 |
| # acquired volumes | 2880 | 312 | 480 |
| recycle delay / s | 5 | 0.2 | 2 |
| measurement time / h | 4 | 50 | 0.3 |
| reconstruction software | Matlab R2018b | ParaVision 6.0.1 | Matlab R2018b |
| reconstructed voxel size / mm$^3$ | 0.15×0.15×1 | 0.09×0.09×0.09 | 0.15×0.15×1 |

## IV. CONCLUSION

In this work, we have taken a crucial step towards model-free investigations of restriction and anisotropy in heterogeneous biological tissues, having potentially far-reaching implications for our understanding of microstructural changes associated with pathology or normal brain development. Importantly, our identification of formal analogies between relaxation and solid-state NMR of rotational dynamics in macromolecules and diffusion MRI of translational motion in biological tissues enabled adaption of existing NMR data acquisition and analysis strategies to the context of microstructural MRI. Through measurements on phantoms, *ex vivo* rat brain, and excised tissue from a mouse model of human neuroblastoma, we demonstrated that our proposed model-free approach is sufficiently flexible to capture the signal modulations for extreme cases of restriction and anisotropy over exhaustive ranges of acquisition variables, while still being robust enough to give quantitative parameter maps reporting on relevant microstructural properties using abbreviated measurement protocols compatible with clinical research studies.

## V. METHODS
### A. Experimental

MRI phantoms with well-defined diffusion properties were assembled from NMR tubes with yeast cell sediment, salt solution, lamellar liquid crystal, and water. Magnesium nitrate hexahydrate, cobalt nitrate hexahydrate, and 1-decanol were purchased from Sigma-Aldrich Sweden AB, sodium octanoate from J&K Scientific via Th. Geyer in Sweden, and fresh baker's yeast (trade name: Kronjäst) at a local supermarket. Unless otherwise stated, water was purified with a Millipore-Q system. The yeast sample was prepared by dispersing a block of yeast in an equal amount of tap water, transferring 1 mL of the cell suspension to a 5 mm NMR tube, and allowing the cells to sediment under the action of gravity at 4 °C overnight [57]. To remove water-soluble nutrients and metabolites contributing to water $T_2$-relaxation via proton chemical exchange, the cells in the tube were washed by three cycles of removing the supernatant with a syringe, adding 2 mL tap water, resuspending by vigorous shaking, and renewed sedimentation at 4 °C. The aqueous salt solution comprised saturated magnesium nitrate [59] doped with cobalt(II) nitrate to reach $T_2$ of about 100 ms. The lamellar liquid crystal was prepared from 85.79 wt% water, 9.17 wt% 1-decanol, and 5.04 wt% sodium octanoate [60]. A composite phantom was assembled by inserting 4 mm NMR tubes with salt solution and liquid crystal into a 10 mm NMR tube with water.

Experiment on *ex vivo* rat brain were approved by the Animal Committee of the Provincial Government of Southern Finland in accordance with the European Union Directives 2010/63/EU. A healthy adult rat Sprague-Dawley was transcardially perfused with 0.9% saline followed by 4% paraformaldehyde in 0.1 M phosphate buffer (pH = 7.4). After extraction, the brain was sagittally sectioned along the brain midline and placed in a solution of phosphate buffered saline 0.1 M and gadoteric acid 50 μl/10 mL (Dotarem 279.3 mg/mL; Guerbet, France) for 24 h. During MRI measurements, the brain was immersed in perfluoropolyether (Galden; TMC Industries, USA) within a 10 mm NMR tube.

Human neuroblastoma cells were cultured at 37° C and 5% $CO_2$ in complete medium (RPMI 1640 supplemented with 10% fetal bovine serum and 1% penicillin/streptomycin). A female BALB/c mouse (Janvier Labs, France) was s.c. inoculated with $2 \cdot 10^6$ tumor cells [34]. After ca. 5 weeks the mouse was sacrificed, the tumor was removed and immediately transferred to a 10 mm NMR tube containing 4% formaldehyde in phosphate buffer solution (Histolab, Sweden). The sample was stored at room temperature for 1 day before being investigated with MRI.

MRI measurements were performed on three different Bruker spectrometers (Karlsruhe, Germany) equipped with MIC-5 probes giving up to 3 $Tm^{-1}$ gradients on-axis. Diffusion encoding employed pairs of variable-angle spinning [30,62] or double rotation [50,51] gradient waveforms bracketing the 180° pulse in a spin echo sequence [71]. Numerical calculation of $\mathbf{b}(\omega)$ included all diffusion and imaging gradients between the centers of the excitation pulse and the spin echo. Additional acquisition and processing parameters are compiled in Table 1. Reconstructed images were exported to NIfTI format for further analysis with the *md-dmri* toolbox [72] in Matlab using custom code based on the novel theory explained in detail below.

### B. Theoretical background

#### 1. Tensor-valued encoding spectrum b(ω)

Within the Gaussian approximation in the cumulant expansion [15,17,48], the diffusion encoding properties are summarized by the tensor-valued encoding spectrum $\mathbf{b}(\omega)$, which is given by the time-dependent magnetic field gradient vector $\mathbf{g}(t)$ via

$$\mathbf{q}(t) = \gamma \int_0^t \mathbf{g}(t')dt', \quad (1)$$

$$\mathbf{q}(\omega) = \int_0^\tau \mathbf{q}(t) \exp(i\omega t) dt, \quad (2)$$

and

$$\mathbf{b}(\omega) = \frac{1}{2\pi} \mathbf{q}(\omega)\mathbf{q}(-\omega)^T. \quad (3)$$

In the equations above, $\gamma$ is the gyromagnetic ratio of the studied atomic nucleus, $\tau$ is the overall duration of the motion-encoding gradients, $\mathbf{q}(t)$ is the time-dependent dephasing vector subject to the echo condition $\mathbf{q}(\tau) = 0$, $\mathbf{q}(\omega)$ is the frequency-domain spectrum of the dephasing vector, and T denotes a matrix transpose. While the full $\omega$-dependent and tensorial representation of $\mathbf{b}(\omega)$ is used in our data processing, we find it instructive to summarize its most important aspects using the magnitude $b$ [68,69], centroid frequency $\omega_{\text{cent}}$ [36], and normalized anisotropy $b_\Delta$ [54]. These variables are defined through the equations

$$b(\omega) = \text{trace}\{\mathbf{b}(\omega)\}, \quad (4)$$

$$\omega_{\text{cent}} = \frac{\int_{-\infty}^{\infty} |\omega| b(\omega) d\omega}{\int_{-\infty}^{\infty} b(\omega) d\omega}, \quad (5)$$

$$\mathbf{b} = \int_{-\infty}^{\infty} \mathbf{b}(\omega) d\omega, \quad (6)$$

$$b = \text{trace}\{\mathbf{b}\} = \int_{-\infty}^{\infty} b(\omega) d\omega, \quad (7)$$

and

$$b_\Delta = \frac{b_A - b_R}{b_A + 2b_R}. \quad (8)$$

In Eq. (6), $\mathbf{b}$ is the conventional ($\omega$-independent) $b$-matrix [68] with axial and radial eigenvalues $b_A$ and $b_R$ and main symmetry axis orientation given by the angles $\Theta$ and $\Phi$ [54].

For a sub-ensemble of spins where the effects of restriction and anisotropy are described with the velocity autocorrelation function $\langle \mathbf{v}(t)\mathbf{v}(t')^T \rangle$ and its Fourier transform, the diffusion spectrum $\mathbf{D}(\omega)$, the signal $S$ at the time $t = \tau$ is given by

$$S = S_0 \exp(-\beta), \quad (9)$$

where $S_0$ is the signal at vanishing gradient amplitude and $\beta$ is the attenuation factor. To prepare for comparisons with the corresponding equations in relaxation and solid-state NMR, the factor $\beta$ is expressed in several equivalent ways that are all familiar from the literature [1,2,48]:

$$\beta = \int_0^\tau \int_0^\tau \mathbf{q}(t)^T \cdot \langle \mathbf{v}(t')\mathbf{v}(t)^T \rangle \cdot \mathbf{q}(t')dt\, dt', \quad (10)$$

$$\beta = \frac{1}{2\pi} \int_{-\infty}^{\infty} \mathbf{q}(\omega)^T \cdot \mathbf{D}(\omega) \cdot \mathbf{q}(-\omega) d\omega, \quad (11)$$

and

$$\beta = \int_{-\infty}^{\infty} \mathbf{b}(\omega) : \mathbf{D}(\omega) d\omega, \quad (12)$$

where the colon denotes a generalized scalar product [69]

$$\mathbf{b}(\omega) : \mathbf{D}(\omega) = \sum_i \sum_j b_{ij}(\omega) D_{ij}(\omega). \quad (13)$$

At each time $t$ or frequency $\omega$, all of $\langle \mathbf{v}(t)\mathbf{v}(t')^T \rangle$, $\mathbf{b}(\omega)$, and $\mathbf{D}(\omega)$ are 3 × 3 symmetric positive-definite matrices with elements $ij \in x,y,z$, while $\mathbf{g}(t)$, $\mathbf{q}(t)$, $\mathbf{q}(\omega)$, and $\mathbf{v}(t)$ are 3 × 1 column vectors with elements $i \in x,y,z$.

#### 2. Formal analogies between relaxation and solid-state NMR and diffusion MRI

The formal analogies with relaxation and solid-state NMR become more apparent when focusing on the special cases of either isotropic restricted ($\omega$-dependent) or anisotropic Gaussian ($\omega$-independent) motion. In the first case, Eqs. (10) and (12) may be written as

$$\beta = \int_0^\tau \int_0^\tau q(t)\langle v(t)v(t')\rangle q(t')\mathrm{d}t\,\mathrm{d}t' \tag{14}$$

and

$$\beta = \int_{-\infty}^{\infty} b(\omega)D(\omega)\mathrm{d}\omega, \tag{15}$$

respectively, where $q(t)$ is the magnitude of $\mathbf{q}(t)$ and $D(\omega)$ is 1/3 of the trace of $\mathbf{D}(\omega)$. Eq. (14) can be recognized as the spin-echo version of the famous Anderson-Weiss model [73] which has been valuable for predicting signal attenuation resulting from molecular reorientation on the time-scales of magic-angle spinning and dipolar recoupling [74]. Here, $\langle v(t)v(t')\rangle$ corresponds to the Anderson-Weiss memory function, closely related to the rotational correlation function in the Lipari-Szabo model-free approach [21], and $q(t)$ is analogous to the function describing the measurement conditions in terms of the timings of radiofrequency pulses and sample spinning. In Eq. (15), $D(\omega)$ takes the role of the spectral density $J(\omega)$, being the Fourier transform of the rotational correlation function, and $b(\omega)$ resembles the set of delta-functions at the frequencies relevant for longitudinal, transverse, or rotating frame relaxation [20]. While relaxation NMR explores the time/frequency-dependence of the rotational correlation function and spectral density by varying the main and radiofrequency magnetic fields, as well as the sample spinning and radiofrequency pulse repetition rates, diffusion MRI relies on spectrally modulated magnetic field gradients [16].

For the anisotropic Gaussian case, Eq. (10) yields [49]

$$\beta = \int_0^\tau q(t)^2 \left(\mathbf{n}(t)^\mathrm{T} \cdot \mathbf{D} \cdot \mathbf{n}(t)\right)\mathrm{d}t, \tag{16}$$

where $\mathbf{n}(t)$ is the unit vector of $\mathbf{q}(t)$ and $\mathbf{D}$ is the plateau value of $\mathbf{D}(\omega)$ at frequencies much lower than any of the characteristic decay rates in $\langle \mathbf{v}(t)\mathbf{v}(t')^\mathrm{T}\rangle$. Comparison with the solid-state NMR equation for the signal evolution during sample reorientation [26] reveals that $q(t)^2$ and $\mathbf{n}(t)$ correspond to the magnitude and direction of the main magnetic field $\mathbf{B}_0$ in the sample-fixed frame, while $\mathbf{D}$ is analogous to the second order tensors describing, for instance, chemical shielding or dipolar couplings. The separation and correlation of the isotropic and anisotropic tensor properties achieved by sample reorientation and radiofrequency pulse sequences in multidimensional solid-state NMR can in diffusion MRI be mimicked by the trajectory of the vector $\mathbf{q}(t)$ [32].

We emphasize that all of the cases in Eqs. (14)-(16) are just different manifestations of the more general expression in Eq. (12). Consequently, although Eqs. (10) and (11) are by far more common in the diffusion literature [1,2], we here favor Eq. (12) for its remarkable versatility, covering both frequency-dependence, corresponding to relaxation NMR, and tensorial aspects, analogous to solid-state NMR, in a surprisingly compact equation that is easily discretized and implemented in data processing code.

*3. $\mathbf{D}(\omega)$-distributions and the Lorentzian approximation*

For a heterogeneous system comprising multiple sub-ensembles with probability given by the distribution $P[\mathbf{D}(\omega)]$, Eq. (9) is here generalized to

$$S[\mathbf{b}(\omega)] = S_0 \int P[\mathbf{D}(\omega)]\exp(-\beta)\mathrm{d}\mathbf{D}(\omega) \tag{17}$$

where $\beta$ is given by the integral of $\mathbf{b}(\omega){:}\mathbf{D}(\omega)$ over $\omega$ according to Eq. (12). For the case of Gaussian ($\omega$-independent) diffusion, Eq. (17) reduces to

$$S(\mathbf{b}) = S_0 \int P(\mathbf{D})\exp(-\mathbf{b}{:}\mathbf{D})\mathrm{d}\mathbf{D}, \tag{18}$$

where $P(\mathbf{D})$ the diffusion tensor distribution that is ubiquitous in diffusion MRI in discrete [27] or continuous [28,29] forms. Here, we set out to characterize restricted and anisotropic diffusion in heterogeneous materials through the function $P[\mathbf{D}(\omega)]$ — the "distribution of tensor-valued diffusion spectra" or, for short, the "$\mathbf{D}(\omega)$-distribution" — by acquiring signals as a function of $\mathbf{b}(\omega)$ and inverting the integral transform in Eq. (17).

To make the data inversion tractable, we make the ansatz that, for each sub-ensemble, $\mathbf{D}(\omega)$ is axially symmetric with frequency-dependent axial and radial eigenvalues, $D_\mathrm{A}(\omega)$ and $D_\mathrm{R}(\omega)$, described with Lorentzian transitions between the zero-frequency values, $D_\mathrm{A}$ and $D_\mathrm{R}$, and the common high-frequency plateau, $D_0$, according to

$$D_{\mathrm{A/R}}(\omega) = D_0 - \frac{D_0 - D_{\mathrm{A/R}}}{1 + \omega^2/\Gamma_{\mathrm{A/R}}^2}, \tag{19}$$

where $\Gamma_\mathrm{A}$ and $\Gamma_\mathrm{R}$ are the frequencies at the centers of the transitions. In the lab frame, $\mathbf{D}(\omega)$ is given by

$$\mathbf{D}(\omega) = \mathbf{R}(\theta,\phi)\cdot \mathbf{D}^{\mathrm{PAS}}(\omega) \cdot \mathbf{R}^{-1}(\theta,\phi), \tag{20}$$

where $\theta$ and $\phi$ are polar and azimuthal angles, $\mathbf{R}(\theta,\phi)$ is a rotation matrix, and

$$\mathbf{D}^{\mathrm{PAS}}(\omega) = \begin{pmatrix} D_\mathrm{R}(\omega) & 0 & 0 \\ 0 & D_\mathrm{R}(\omega) & 0 \\ 0 & 0 & D_\mathrm{A}(\omega) \end{pmatrix} \tag{21}$$

is the diffusion spectrum in its principal axis system (PAS). With this approximation, each discrete component in the $\mathbf{D}(\omega)$-distribution can be described with its statistical weight $w$ and the parameter set $[D_\mathrm{A},D_\mathrm{R},\theta,\phi,D_0,\Gamma_\mathrm{A},\Gamma_\mathrm{R}]$.

The functional form of Eq. (19) can be justified by comparison with the corresponding multi-Lorentzian expression for a liquid undergoing restricted diffusion along the principal axes of planar, cylindrical, and spherical compartments [17,48]:

$$D(\omega) = \frac{D_0}{1 + \omega^2/\Gamma_0^2} - \sum_k w_k \frac{D_0 - D_\infty}{1 + \omega^2/\Gamma_k^2}. \tag{22}$$

This version of the well-known equation includes effects of the molecular-level transition from ballistic to diffusive motion via the decay rate $\Gamma_0$ of the assumedly exponential velocity autocorrelation function [14], as well as the compartment-level transition between the bulk and long-range diffusivities $D_0$ and $D_\infty$, the latter taking the finite permeability of the compartment walls into account [75]. The transition is determined by the weights $w_k$ and Lorentzian widths $\Gamma_k$ given by

$$w_k = \frac{2}{\zeta_k^2 + 1 - d} \quad (23)$$

and

$$\Gamma_k = \frac{\zeta_k^2 D_0}{r^2}, \quad (24)$$

where $d = 1$, 2, and 3 for, respectively, the planar, cylindrical, and spherical cases, $r$ is the compartment radius, $\zeta_k$ is the $k$th solution of

$$\zeta J_{d/2-1}(\zeta) - (d-1)J_{d/2}(\zeta) = 0, \quad (25)$$

and $J_\nu$ is the $\nu$th order Bessel function of the first kind. The sum in Eq. (22) is dominated by the first term and the sum of all $w_k$ equals unity, indicating that the multi-Lorentzian expression can be approximated with a single Lorentzian as in Eq. (19). Computer simulations of water at 298 K show that simple exponential autocorrelation is a rather crude approximation [63], but for the purpose of this paper it is sufficient to note that $\Gamma_0$ is on the order of $10^{13}$ s$^{-1}$ and, within the regime accessible with NMR methods based on magnetic field gradients, the first term of Eq. (22) can be approximated with $D_0$ as in Eq. (19).

### C. Monte Carlo inversion and extraction of relevant metrics

For each set of $\mathbf{b}(\omega)$-encoded signals, ensembles of discrete $\mathbf{D}(\omega)$-distributions are estimated by Monte Carlo inversion [31] that has previously been applied and described in detail for various diffusion and relaxation correlation measurements including $[D_A,D_R]$ [32], $[D_A,D_R,\theta,\phi]$ [33,64], and $[D_A,D_R,\theta,\phi,R_1,R_2]$ [55]. In terms of data inversion, the extension to the $[D_A,D_R,\theta,\phi,D_0,\Gamma_A,\Gamma_R]$-space is straightforward, and for information on the algorithm we refer the reader to the previous literature [55] and the corresponding Matlab code available at https://github.com/daniel-topgaard/md-dmri. Following the terminology in previous papers, the inversion was here performed with the limits $5 \cdot 10^{-12}$ m$^2$s$^{-1} < D_{0/A/R} < 5 \cdot 10^{-9}$ m$^2$s$^{-1}$ and 0.1 s$^{-1} < \Gamma_{A/R} < 10^5$ s$^{-1}$, 20 steps of proliferation, 20 steps of mutation/extinction, 200 input components per steps of proliferation and mutation/extinction, 10 output components, and bootstrapping by 100 repetitions using random sampling with replacement.

While the individual realizations of the ensemble of solutions are "overfits"—containing spurious details consistent with, but not necessarily required by, the acquired data—it is possible to derive coarse-grained metrics, such as means and (co)variances over relevant dimensions, that are determined with higher precision quantifiable via bootstrapping [33]. Parameters values shown as projections and maps in the figures are obtained by taking the medians of the individual values for the ensemble of solutions. In the current context, the characteristic frequencies $\Gamma_{A/R}$ require special attention as they assume a role equivalent to the rotational correlation times in the interpretation of NMR relaxation dispersion data [20]. Acknowledging that the allowed values of $\Gamma_{A/R}$ extend beyond the range of $\omega$ actually encoded in the data, we use insights from the recent concept of dynamics detectors in relaxation NMR [22-24] to convert the noisy ensembles of $\mathbf{D}(\omega)$-distributions in the primary analysis space $[D_A,D_R,\theta,\phi,D_0,\Gamma_A,\Gamma_R]$ into quantities better supported by the data. In practice, this means evaluating the $\mathbf{D}(\omega)$-distributions at selected values of $\omega$ within the narrow range probed by the gradient waveforms, giving $[D_A(\omega),D_R(\omega),\theta,\phi]$ via Eq. (19), projecting onto the dimensions of isotropic diffusivity $D_{iso}(\omega)$ and squared normalized anisotropy $D_\Delta(\omega)^2$ [56] through

$$D_{iso}(\omega) = \frac{D_A(\omega) + 2D_R(\omega)}{3} \quad (26)$$

and

$$D_\Delta(\omega)^2 = \frac{[D_A(\omega) - D_R(\omega)]^2}{[D_A(\omega) + 2D_R(\omega)]^2}, \quad (27)$$

and calculating per-voxel and bin-resolved means E[$x$], variances V[$x$], and covariances C[$x,y$] over the diffusion dimensions [33]. Following previous works using oscillating gradient encoding [36,65,67], the effects of restriction are quantified as a finite difference approximation of the rates of change of these metrics within the investigated frequency window, for instance

$$\Delta_{\omega/2\pi}\mathrm{E}[D_{iso}] = \frac{\mathrm{E}[D_{iso}(\omega_{max})] - \mathrm{E}[D_{iso}(\omega_{min})]}{(\omega_{max} - \omega_{min})/2\pi}. \quad (28)$$

### ACKNOWLEDGMENTS

This work was financially supported the Swedish Foundation for Strategic Research (ITM17-0267), Swedish Research Council (2018-03697, 21073), Swedish Cancer Society (3427), Swedish Childhood Cancer Fund, Academy of Finland (#323385), and Erkko Foundation. The Swedish NMR Centre in Gothenburg is acknowledged for spectrometer time.

### COMPETING INTERESTS

D.T. owns shares in Random Walk Imaging AB (Lund, Sweden, http://www.rwi.se/), holding patents related to the described methods.

### DATA AND CODE AVAILABILITY

Upon manuscript acceptance, data and code will be made available at https://github.com/daniel-topgaard/.

### AUTHOR CONTRIBUTIONS

O.N and A.S: preparation of rat brain; acquisition, processing, and interpretation of MRI data. M.Y.: development of ParaVision data acquisition code and Matlab conversion code; acquisition of MRI data. H.J.: development and preparation of phantoms; acquisition and processing of MRI data. E.F.-A.: development of tumor model, interpretation of tumor data. D.B.: preparation of tumor samples